\documentclass[aps,prb,reprint,superscriptaddress]{revtex4-1}

\usepackage{amsmath}    
\usepackage{graphicx}   
\usepackage[lofdepth,lotdepth, caption=false]{subfig}
\usepackage{verbatim}   
\usepackage{color}      
\usepackage{hyperref}   
\usepackage{dcolumn}

\begin{document}
\newcommand{\sninte}{Sn$_{1-x}$In$_{x}$Te}
\newcommand{\degrees}{$^\circ$C}
\newcommand{\Tc}{$T_{c}$}
\newcommand{\at}[2][]{#1|_{#2}}


\title{Optimizing the superconducting transition temperature and upper critical field of \sninte}
\author{R.~D.~Zhong}

\affiliation{Condensed Matter Physics and Materials Science Department, Brookhaven National Laboratory, Upton, NY 11973, USA}
\affiliation{Materials Science and Engineering Department, Stony Brook University, Stony Brook, NY 11794, USA}
\author{J.~A.~Schneeloch}
\affiliation{Condensed Matter Physics and Materials Science Department, Brookhaven National Laboratory, Upton, NY 11973, USA}
\affiliation{Department of Physics and Astronomy, Stony Brook University, Stony Brook, NY 11794, USA}
\author{X.~Y.~Shi}
\affiliation{Condensed Matter Physics and Materials Science Department, Brookhaven National Laboratory, Upton, NY 11973, USA}
\author{Z.~J.~Xu}
\affiliation{Condensed Matter Physics and Materials Science Department, Brookhaven National Laboratory, Upton, NY 11973, USA}
\author{C.~Zhang}
\affiliation{Condensed Matter Physics and Materials Science Department, Brookhaven National Laboratory, Upton, NY 11973, USA}
\affiliation{Materials Science and Engineering Department, Stony Brook University, Stony Brook, NY 11794, USA}
\author{J.~M.~Tranquada}
\author{Q.~Li}
\author{G.~D.~Gu}
\affiliation{Condensed Matter Physics and Materials Science Department, Brookhaven National Laboratory, Upton, NY 11973, USA}

\date{\today} 

\begin{abstract}
\sninte\ is a possible candidate for topological superconductivity.  Previous work has shown that substitution of In for Sn in the topological crystalline insulator SnTe results in superconductivity, with the transition temperature, \Tc, growing with In concentration.  We have performed a systematic investigation of \sninte\ for a broad range of $x$, synthesizing single crystals (by a modified floating zone method) as well as polycrystalline samples.  The samples have been characterized by x-ray diffraction, resistivity, and magnetization.  For the single crystals, the maximum \Tc\ is obtained at $x=0.45$ with a value of 4.5~K, as determined by the onset of diamagnetism. 
\end{abstract}


\maketitle

Topological crystalline insulators (TCIs),\cite{Fu_Topological_2011} in which the metallic surface states are protected by crystal point-group symmetry instead of the time-reversal symmetry that contributes to the traditional topological insulators (TIs),\cite{hasa10,qi11} 
have attracted great interest since their discovery. The theoretical prediction\cite{Hsieh_topological_2012} of a TCI phase in SnTe has received strong experimental support from angle-resolved photoemission studies.\cite{Tanaka_experimental_2012,Xu_topological_2012} This, in turn, has triggered the search for possible topological  superconductivity\cite{Fu_superconducting_2008,qi11} in the SnTe system.  In particular, spectroscopic studies\cite{sasaki_odd-parity_2012,sato_2013} have recently examined In-doped SnTe, as it is a low-carrier-density superconductor based on a narrow-gap semiconductor, which satisfies the criteria of possible topological superconductivity.\cite{sasaki_odd-parity_2012,Hsieh_Majorana_2012,Fu_odd-parity_2010} Earlier studies of superconductivity in \sninte\ have shown that one can vary $T_c$ from $<1$~K at $x=0.02$ to 2.6~K at $x=0.2$.\cite{Bushmarina_superconducting_1991,Parfeniev_superconductivity_1995,Erickson_enhanced_2009,Erickson_anomalous_2010}  That work suggests that one should be able to raise $T_c$ by further increasing the In concentration.

\begin{figure}[b]
\begin{center}
\includegraphics[width=8.6cm]
{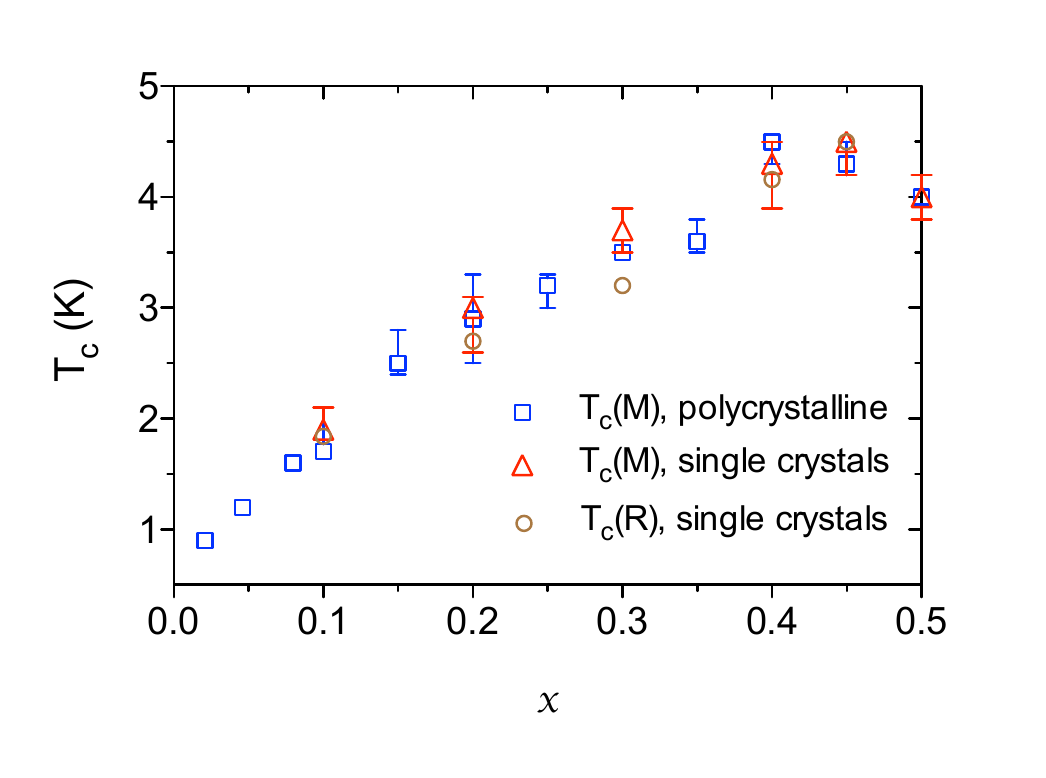}
\caption{\label{fig:PhaseDiagram} (Color online) Superconducting transition temperature as a function of indium concentration $x$ for both polycrystalline (blue squares) and single crystal samples (red triangles), obtained from magnetization measurements, and for single crystals (brown circles) from resistivity measurements.  For each concentration, different parts of the as-grown crystal rod were measured to give an average value of \Tc.  Data for indium concentrations less than 10\% are taken from Erickson {\it et al.}\cite{Erickson_enhanced_2009} 
}
\end{center}
\end{figure}

Given the interest in the \sninte\ system, we decided to perform a systematic study of $T_c$ and the upper critical field, $H_{c2}$, as a function of $x$.  A series of both polycrystalline and single-crystal samples 
have been synthesized and characterized.  Exploring the composition range $0\le x\le 1$, single-phase samples were obtained for $x\lesssim0.5$.  The results for $T_c$ are summarized in Fig.~\ref{fig:PhaseDiagram}.  For the single crystals, where the composition is best controlled, we find that $T_c$ has a maximum of 4.5~K at $x=0.45$, with an estimated $\mu_{0} H_{c2}(T=0) = 1.52$~T.  For polycrystalline samples, we found the same maximum $T_c$ at a nominal $x=0.40$.  Following an initial presentation of our results,\cite{zhong13} we became aware of the work of Balakrishnan {\it et al.},\cite{Balakrishnan_superconducting_2013} who confirmed the magnitude of $T_c$ at $x=0.4$.\footnote{The reported onset temperature of superconductivity in Ref.~\onlinecite{Balakrishnan_superconducting_2013} is similar to our result, but the resistive transition is much broader than that observed for our single crystals.}

Polycrystalline samples with nominal composition \sninte\ $(0\leq x\leq1.0)$ were prepared via the horizontal unidirectional solidification method. Stoichiometric mixtures of high purity elements [Sn (99.99\%), In (99.99\%) and Te (99.999\%)] were sealed under vacuum in double-walled quartz ampoules. The ampoules were heated at 850\degrees\ in a box furnace and rocked to achieve good mixing of the ingredients. Afterward, the samples were gradually cooled down to room temperature over 24 hours and removed for characterization. Single crystals with nominal In concentration of $x=0.1$--0.5 were grown by a modified floating zone method.  The starting material was prepared as for the polycrystalline samples, after which the quartz ampoule was mounted in a floating-zone furnace.  The space around the quartz was filled with high-purity Ar at 1 bar, to avoid oxygen diffusion through the quartz.  A crystal growth velocity of 0.5--1~mm/h was used.      

The crystal structure of each composition was characterized by X-ray powder diffraction measured with Cu $K\alpha$ radiation at room temperature. The diffraction data were analyzed with the GSAS program package.\cite{GSAS} For each single-crystal composition, a crushed piece was used for the diffraction measurement.  Magnetic measurements were carried out in a Quantum Design superconducting quantum interference device (SQUID) magnetometer, for temperatures down to 1.75~K. Electrical resistivity was measured using the in-line four-point configuration, with an excitation current of 5~mA, in a Quantum Design Physical Properties Measurement System (PPMS).

\begin{figure}[t]
\begin{center}
\includegraphics[width=8.5cm]
{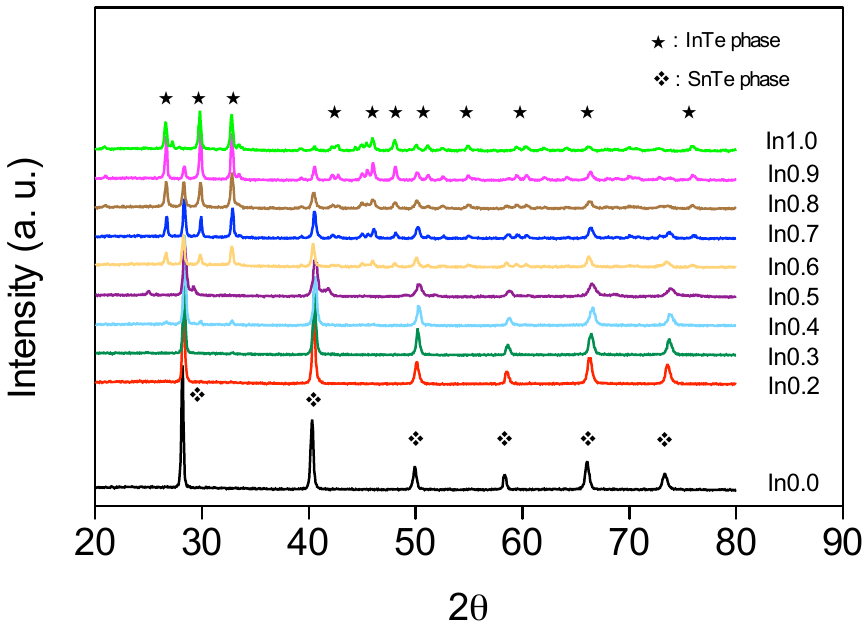}
\caption{\label{fig:XRDPolycrystalline} (Color online) X-ray diffraction spectra for \sninte\ (In$x$) polycrystalline samples with nominal compositions $x=0$--1.0.}
\end{center}
\end{figure}

\begin{figure}[t]
\begin{center}
\includegraphics[width=8.5cm]
{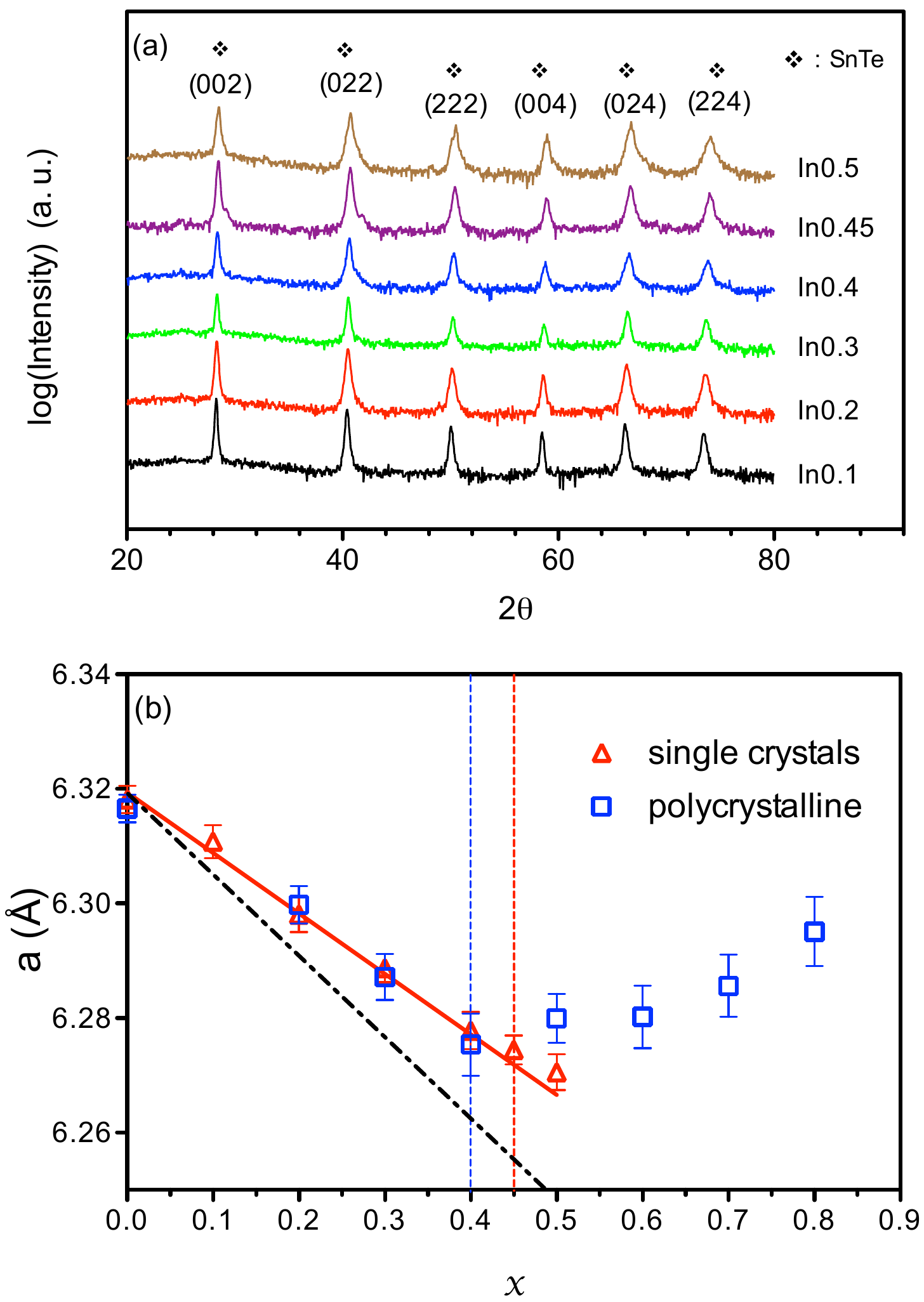}
\caption{\label{fig:XRDsinglecrystal} (Color online) (a) X-ray diffraction measurements for the single crystals with composition $x=0.1$--0.5, with intensity plotted on a logarithmic scale. (b) Lattice parameters $a$ derived from Rietveld analysis of XRD patterns of the cubic phases in polycrystalline (blue squares) and single crystals (red triangles) samples as a function of the nominal indium concentration $x$.}
\end{center}
\end{figure}

Each polycrystalline sample rod was approximately 10 cm in length.  A piece was cut from the center for characterization.  Each crystal rod grown by the modified floating-zone method was approximately 15 cm in length.  Pieces cut from the center and ends were characterized by magnetization measurements.  The error bars on $T_c$ in Fig.~\ref{fig:PhaseDiagram} reflect the observed spread in values and provide an indirect measure of compositional variation.  For the x-ray diffraction and resistivity measurements, a piece from the center of each crystal rod was used.

X-ray diffraction (XRD) was used to investigate the crystal structure. As shown in Fig.~\ref{fig:XRDPolycrystalline}, the peaks of the indium-free sample (SnTe) can be indexed quite well to the rocksalt structure with a lattice constant of $a=6.318$~\AA, consistent with previous studies.\cite{sasaki_odd-parity_2012,Hsieh_topological_2012}  Similarly, XRD patterns were obtained for each of the polycrystalline samples ($x=0.1$--1.0) as shown in Fig.~\ref{fig:XRDPolycrystalline}.  Only the cubic phase is detected for $x\le0.3$, while the $x=0.4$ sample shows a trace amount ($\lesssim2$\%) of a secondary phase (tetragonal InTe).  For $x=0.5$, an undetermined secondary phase is present, while the tetragonal InTe phase becomes substantial for $x\ge0.6$.

The XRD patterns for the single-crystal growths are plotted on a logarithmic scale in Fig.~\ref{fig:XRDsinglecrystal}(a).  There is no detectable second phase for $x\le0.4$.  For $x=0.45$, there is a small second phase, while $x=0.5$ exhibits asymmetrically broadened diffraction peaks, but no obvious second phase.

The lattice parameter $a$ determined for the cubic phase in the polycrystalline and single-crystal samples are illustrated in Fig.~\ref{fig:XRDsinglecrystal}(b).  According to Vegard's law, the lattice parameters of the alloy samples should vary linearly between the values of the end members.  There is a complication in the present case in that the stable phase of InTe is tetragonal.  The cubic phase is stable at high pressure, and it can be retained as a metastable phase at ambient pressure with reported $a=6.177$~\AA\ at room temperature.\cite{darnell_1964}   The corresponding Vegard's law prediction is given by the dot-dashed line, which we present for completeness; however, the relevance of this line is unclear given the instability of the InTe end point.

We find that the lattice parameters of the single-crystal samples follow a straight line but with a slightly reduced slope.  The polycrystalline samples follow the same line up to $x=0.4$, but deviate from it for $x\ge0.5$, where substantial amounts of second phase are evident.  Clearly, the In concentration in the cubic-phase alloy reaches its maximum at $x\sim0.5$, and that saturation leads to the presence of the InTe second phase for larger $x$.  Given the linear evolution of $a$ and the single-phase behavior for $x<0.5$, we conclude that the actual In concentration is approximately equal to the nominal concentration in cubic phase Sn$_{1-x}$In$_x$Te for $x<0.5$.

\begin{figure}[b]
\begin{center}
\includegraphics[width=8.6cm]
{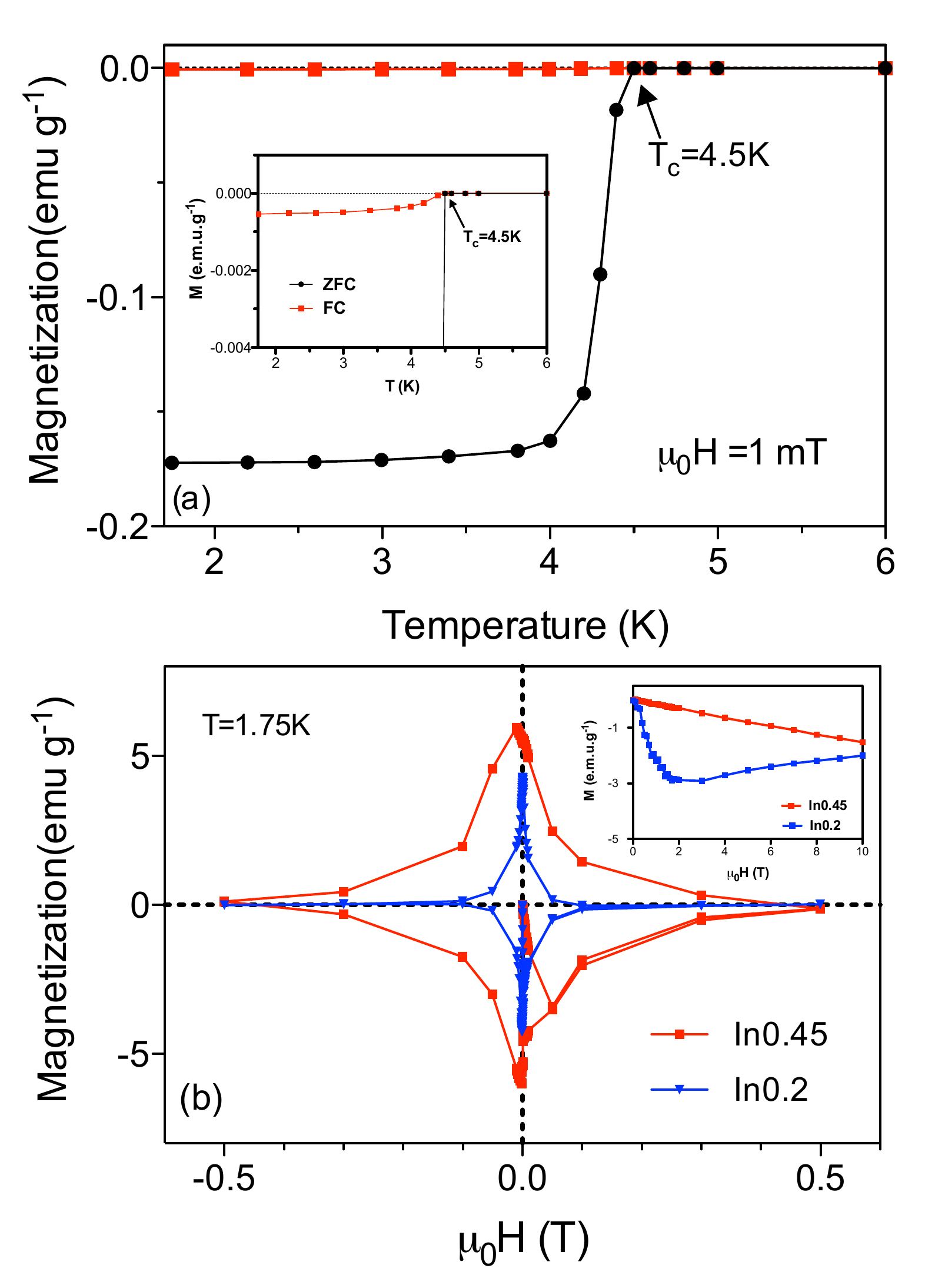}
\caption{\label{fig:Magnetization} (Color online) (a) Temperature dependence of magnetic susceptibility for Sn$_{0.55}$In$_{0.45}$Te single crystals measured under conditions of FC (field cooled) and ZFC (zero-field cooled) in an applied field of 1~mT at a cooling/heating rate of 0.1 K/min. (b) Hysteresis loop of a Sn$_{0.55}$In$_{0.45}$Te single crystal measured at 1.75~K. The inset shows the initial M-H behavior at field less than 10~mT and 1.75~K. For magnetization measurements, samples are cut into roughly cubic-like small chunks. The demagnetization factor can be taken as 1 considering that the crystal structure is isotropic. }
\end{center}
\end{figure}

The superconducting transition temperature of each sample was determined by magnetization measurements.  Figure~\ref{fig:Magnetization}(a) shows the temperature dependence of the zero-field cooled (ZFC) and field cooled (FC) magnetizations for the Sn$_{0.55}$In$_{0.45}$Te single crystal measured between 1.75~K and 6~K under an applied field of 1~mT. The onset of the Meissner signal occurs at $T_c=4.5$~K. The magnetic hysteresis ($M$-$H$) loops for representative $x=0.2$ and $x=0.45$ single crystal samples measured at 1.75~K are compared in Fig.~\ref{fig:Magnetization}(b). Clearly the $M$-$H$ loop for In$0.45$ includes a larger area than that of In$0.2$, indicating stronger flux pinning for 45\%\ indium substitution. The lower critical field $H_{c1}$,  defined by the deviation point of the $M$-$H$ curve from its initial linear behavior, can be estimated from the data in the inset of Fig.~\ref{fig:Magnetization}(b).  With higher indium concentration, \sninte\ superconductors are more likely to resist magnetic flux penetration. 

\begin{figure}[t]
\begin{center}
\includegraphics[width=7.5cm]
{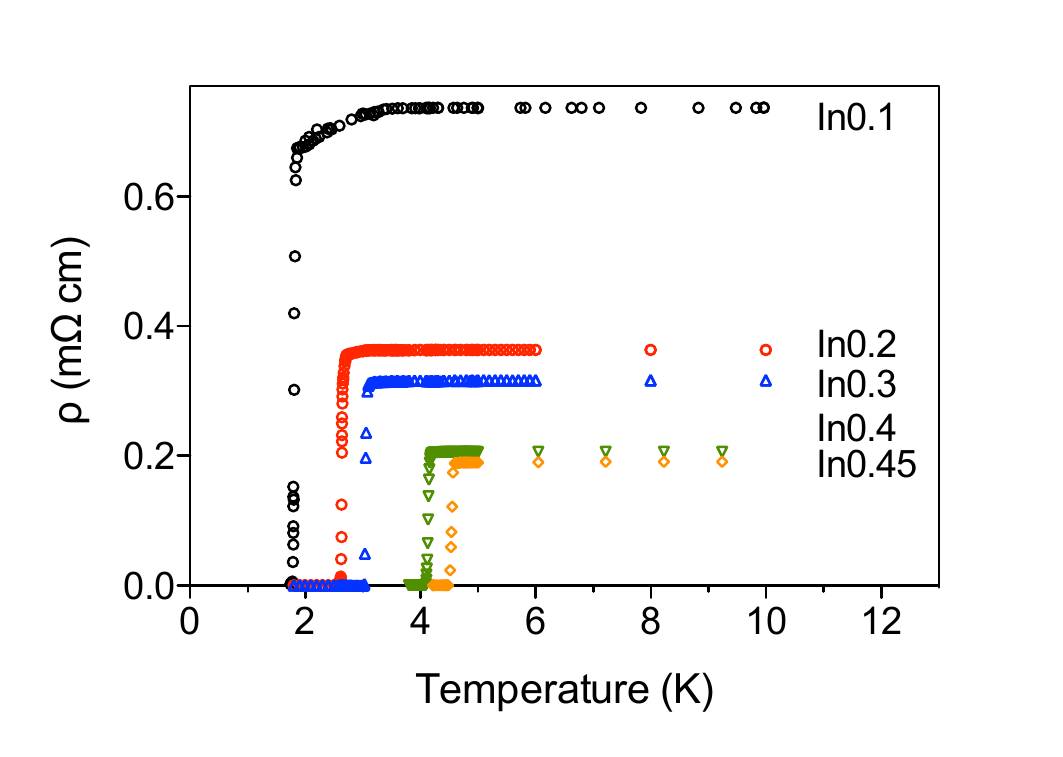}
\caption{\label{fig:Resistivity} (Color online) Temperature dependence of the resistivity for \sninte\ single crystals.}
\end{center}
\end{figure}

\begin{figure}[t]
\begin{center}
\includegraphics[width=7.5cm]{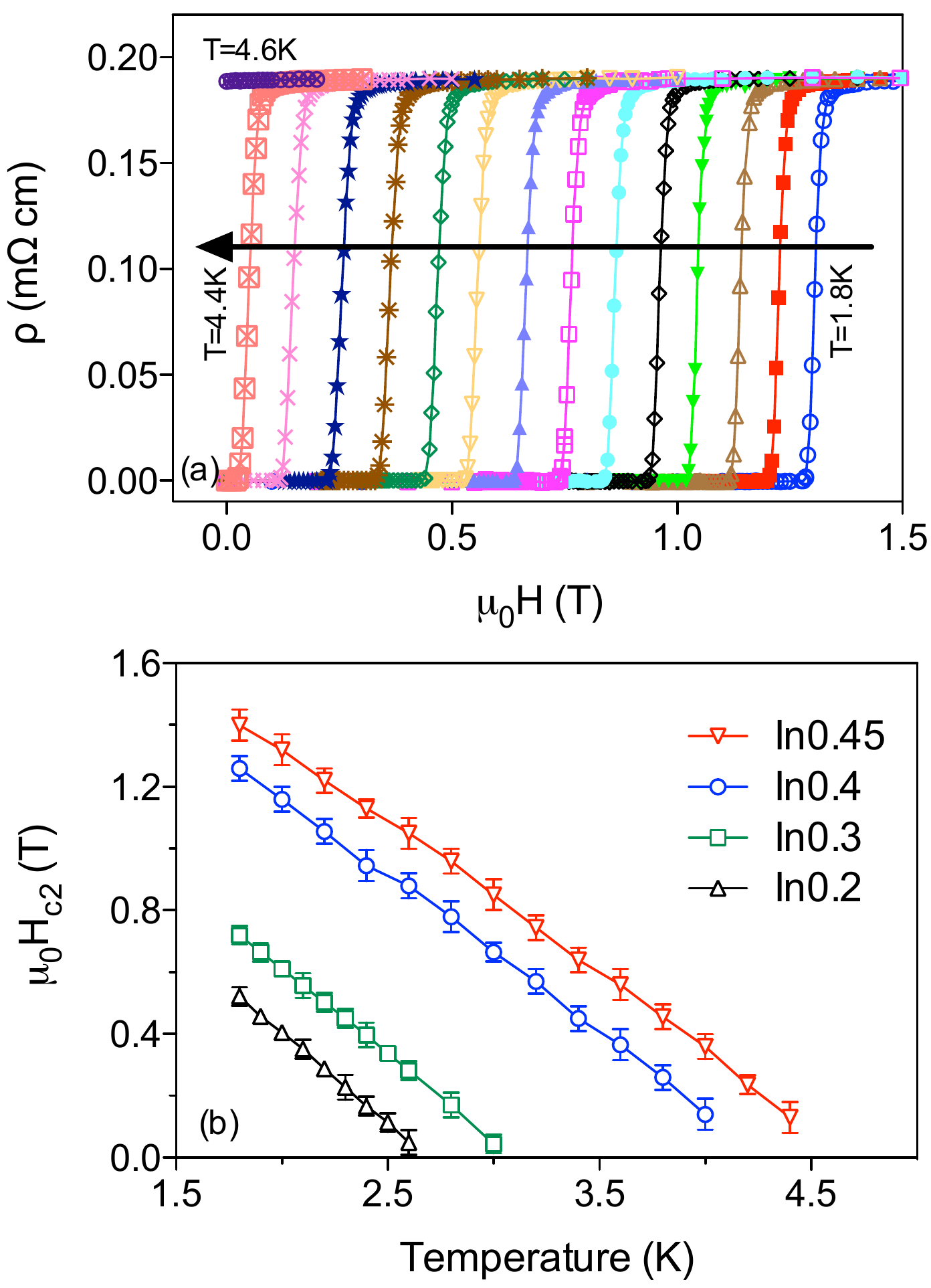}
\caption{\label{fig:Transport} (Color online) (a) Field dependence of the resistivity for the $x=0.45$ crystal at  fixed temperatures from 1.8 to 4.6~K. (b) Upper critical field H$_{c2}(T)$ determined from resistivity measurements on four single crystals.}
\end{center}
\end{figure}

The superconducting critical temperature $T_c(M)$ is defined as the temperature at which the magnetic moment begins to drop sharply.  The results for all samples are displayed in Fig.~\ref{fig:PhaseDiagram}.  Starting at small $x$, the polycrystalline samples show an almost linear relation between $T_c$ and $x$, reaching the highest \Tc\ of 4.5~K when $x=0.4$.\footnote{The growth technique for the polycrystalline samples results in some degree of variation in In concentration along each sample. We have thoroughly discussed the relation between the actual concentration and nominal concentration based on the XRD results. As a consequence, we consider the nominal $x$ values of the polycrystalline samples to be less reliable than those of the single crystals.  The spread in In concentration in the polycrystalline samples is indicated by significantly broader superconducting transitions, measured by magnetization, than those measured for single crystals.} The variation of $T_c$ with $x$ for single crystals shows a similar relation for $x\le0.45$.  This trend confirms and extends the earlier experimental results from several groups,\cite{Bushmarina_superconducting_1991,Parfeniev_superconductivity_1995,Erickson_enhanced_2009,Erickson_anomalous_2010} and it occurs within the regime where the XRD results indicate that the samples are essentially single phase.  

The resistivity of \sninte\ single crystal samples is weakly metallic in the normal state up to 300~K (not shown); the transition to zero resistance is fairly sharp, with a typical width of 0.2~K, as shown in Fig.~\ref{fig:Resistivity}. The parameter $T_{c}(R)$ is defined as the onset temperature for the drop in resistivity. As the indium concentration increases, $T_{c}(R)$ changes similarly with $T_{c}(M)$, as shown in Fig.~\ref{fig:PhaseDiagram}.  

The magnetic-field dependence of the electrical resistivity for single crystals was also measured. Representative data for Sn$_{0.55}$In$_{0.45}$Te is shown in Fig.~\ref{fig:Transport}(a). The upper critical field $H_{c2}$ was determined from such measurements. Specifically, $H_{c2}(T)$ is defined as the onset of the resistive transition at each fixed temperature; the results are shown in Fig.~\ref{fig:Transport}(b).  The upper critical field at zero temperature $H_{c2}(T=0)$ can be estimated using the Werthamer-Helfand-Hohenberg approximation,\cite{Werthamer_Temperature_1966} $H_{c2}(0)=0.69 T_c |dH_{c2}/dT|_{T=Tc}$.  From the measured curves for $\mu_{0} H_{c2}(T)$, we obtain the initial slopes (at $T\approx T_c$) of $-0.58$~T/K for $x=0.2$, $-0.55$~T/K for $x=0.3$, $-0.49$~T/K for $x=0.4$, and $-0.48$~T/K for $x=0.45$. Using $T_{c}=2.7$~K, 3.1~K, 4.2~K and 4.5~K for the respective samples, the upper critical fields are estimated to be $\mu_{0} H_{c2}(T=0)=1.08$~T,  1.18~T, 1.42~T, and 1.49~T for $x=0.2$, 0.3, 0.4, and 0.45, respectively. These results show that stronger applied field is needed to completely suppress the superconductivity at 0~K for \sninte\ superconductors with higher indium concentrations.

To conclude, we have investigated the correlations between indium concentration, crystal structure, and superconducting properties for the In-doped SnTe system. Bulk \sninte\ single crystals were successfully grown by the modified floating zone method. In our work we have demonstrated that the material retains a single phase of the rocksalt structure with linearly shrinking lattice parameters for $x\le0.4$ in polycrystalline samples and $x\le0.45$ in single crystals. With indium substitution, \sninte\ displays bulk superconductivity at accessible temperatures, and indium shows a significant effect on enhancing the \Tc\ of \sninte\ materials. A maximum in \Tc\ around 4.5~K was achieved in both polycrystalline and single crystals at $x=0.4$ and $x=0.45$, respectively.   Transport measurements show that raising the indium concentration towards the optimum value also leads to a higher upper critical field. For Sn$_{0.55}$In$_{0.45}$Te single crystals, the $\mu_{0} H_{c2}(T=0)$ is estimated to be 1.49~T.

Work at Brookhaven is supported by the Office of Basic Energy Sciences, Division of Materials Sciences and Engineering, U.S. Department of Energy under contract No.\ DE-AC02-98CH10886. The synthesis, magnetization, and x-ray diffraction measurements were supported by the Center for Emergent Superconductivity, an Energy Frontier Research Center, while the resistivity measurements were supported by the Superconducting Materials project.


%

\end{document}